\documentclass[10pt,twocolumn,letterpaper]{article}

\usepackage{cvpr}              

\usepackage[dvipsnames]{xcolor}

\usepackage{graphicx}
\usepackage[export]{adjustbox}
	
\usepackage[accsupp]{axessibility}  

\definecolor{cvprblue}{rgb}{0.21,0.49,0.74}
\usepackage[pagebackref,breaklinks,colorlinks,citecolor=cvprblue]{hyperref}

\def\paperID{30} 
\def\confName{CVPR}
\def\confYear{2024}

\title{Unimodal Multi-Task Fusion for Emotional Mimicry Intensity Prediction}

\author{Tobias Hallmen\\
Chair for Human-Centered Artificial Intelligence\\
University of Augsburg\\
{\tt\small tobias.hallmen@uni-a.de}
\and
Fabian Deuser\\
Institute for Distributed Intelligent Systems\\
University of the Bundeswehr Munich\\
{\tt\small fabian.deuser@unibw.de}
\and
Norbert Oswald\\
Institute for Distributed Intelligent Systems\\
University of the Bundeswehr Munich\\
{\tt\small norbert.oswald@unibw.de}
\and
Elisabeth André\\
Chair for Human-Centered Artificial Intelligence\\
University of Augsburg\\
{\tt\small elisabeth.andre@uni-a.de}
}

\begin{document}
\maketitle






\begin{abstract}

In this research, we introduce a novel methodology for assessing Emotional Mimicry Intensity (EMI) as part of the 6th Workshop and Competition on Affective Behavior Analysis in-the-wild. Our methodology utilises the Wav2Vec 2.0 architecture, which has been pre-trained on an extensive podcast dataset, to capture a wide array of audio features that include both linguistic and paralinguistic components. We refine our feature extraction process by employing a fusion technique that combines individual features with a global mean vector, thereby embedding a broader contextual understanding into our analysis. 
A key aspect of our approach is the multi-task fusion strategy that not 
only leverages these features but also incorporates a pre-trained 
Valence-Arousal-Dominance (VAD) model. This integration is designed to 
refine emotion intensity prediction by concurrently processing multiple emotional dimensions, thereby embedding a richer contextual understanding into our framework.
For the temporal analysis of audio data, our feature fusion process utilises a Long Short-Term Memory (LSTM) network. This approach, which relies solely on the provided audio data, shows marked advancements over the existing baseline, offering a more comprehensive understanding of emotional mimicry in naturalistic settings, achieving the second place in the EMI challenge.
\end{abstract}

\section{Introduction}
Emotional mimicry is a phenomenon where individuals imitate the emotional expressions of others, such as their facial expressions, vocal tones, or body language~\cite{hess1999mimicry, hess2013emotional, hess2022theoretical}. This mirroring can facilitate social bonding by helping individuals understand and empathise with the emotional states of those around them~\cite{hess2022theoretical}. For example, when someone smiles, others around them are likely to do the same, creating a shared emotional experience. One particular interesting aspect that this mimicry phenomenon mostly appears when people already have bonds to each other~\cite{likowski2008modulation}. In therapeutic settings, emotional mimicry can be particularly beneficial, as it helps therapists connect with their clients, making them feel understood and supported. 

Leveraging deep learning to discern and anticipate emotional states in therapeutic settings enhances therapists' insight into and reaction to clients' emotions. Utilising indicators like facial expressions, and variations in voice pitch and tone aids in this process. Previous research~\cite{christ2023muse} introduced a multi-modal dataset categorising emotional mimicry into ``Approval'', ``Disappointment'', and ``Uncertainty''. Contemporary studies~\cite{guofeng2023llm, groz2023electra, ding2023multimodal} employ diverse multimodal features from this dataset, including audio-visual and, with the aid of ASR models~\cite{radford2023robust}, textual data, further enhanced by textual embeddings~\cite{zeng2022glm, devlin2018bert, clark2020electra}. Integrating large language model features with audio-visual data for predicting emotional mimicry brings considerable computational challenges. This is due to the intricate process of combining and analysing different types of data. Therefore, creating a cohesive end-to-end pipeline becomes an important goal. Such a pipeline would provide a harmonious balance between efficiency and performance when handling diverse data streams.

For this study, we participated in the 6th Workshop and Competition on Affective Behavior Analysis in-the-wild, specifically the Emotional Mimicry Challenge. This segment of the challenges focuses on predicting the intensity of mimicked emotions, a crucial aspect that could enhance the precision of therapeutic applications. The challenge utilises video data showcasing individuals mimicking specific seed emotions, with a notable twist: the annotators were unaware of the intended emotions of the seeds, providing intensity ratings based on their mimicry and seeded emotions. This approach allowed for a more authentic assessment of emotional mimicry intensity, offering valuable insights into how emotions are conveyed and perceived in a naturalistic setting. Compared to the previous dataset~\cite{christ2023muse} it contains a more fine-grained categorisation of the emotions, which is more challenging to separate.

We present an innovative method focused solely on audio to predict emotional mimicry in response to perceived videos. This approach stems from our preliminary findings, which indicated that relying on visual data did not yield satisfactory outcomes and imposed significant hardware limitations during the training process. By exclusively utilising audio data, we conducted evaluations of the given challenge dataset and achieved competitive results. Our findings underscore the critical role of task-specific pre-trained weights and highlight the significance of our architectural decisions in the success of this method.

\section{Related work}
Within the domain of affective computing, notable advancements have been achieved in a variety of tasks, each contributing to an enhanced comprehension of emotional expressions and their computational detection. The scope of these tasks encompasses the recognition of emotional expressions~\cite{kollias2019expression, kollias2020analysing}, the detection of facial action units~\cite{kollias2019face, kollias2020analysing}, regression analysis of valence and arousal~\cite{zafeiriou2017aff,kollias2019deep,kollias2020analysing,kollias2021affect}, and, more recently, the estimation of emotional mimicry~\cite{christ2023muse}. These tasks are systematically organised within the Affective Behavior Analysis in-the-wild (ABAW) Challenge~\cite{kollias2022abaw,kollias2023abaw,kollias2023abaw2,kollias2023multi}, which, in its latest iteration~\cite{kollias20246th}, has introduced emotional mimicry in a more nuanced way as a novel area of exploration, reflecting the field's evolving focus and expanding methodologies.

Gr\'{o}sz et al.~\cite{groz2023electra} present an multimodal approach on the MuSe Mimic dataset by integrating audio, visual, and textual data. They utilised fine-tuned Wav2vec 2.0 features to analyse audio inputs, extracted facial action units from video data to interpret visual cues, and employed an ELECTRA text encoder~\cite{clark2020electra} for textual feature extraction. These diverse data streams were then synergistically combined using a late fusion technique.

In contrast to a conventional late fusion method, Ding et al.~\cite{ding2023multimodal} implement a specialised fusion block with cross-attention mechanisms for the integration of multimodal data. This novel component is specifically trained to manage the fusion of different modalities, enhancing the model's capability to identify key features across the varied data types. This method represents a more nuanced approach to feature integration, aiming to improve the overall efficacy.

Adopting a multimodal transformer with cross-attention for data fusion, Guofeng et al.~\cite{guofeng2023llm} distinguishes itself by utilising features from a large language model rather than conventional text encoders. This approach provides a more nuanced understanding of text, enriching the model's interpretative depth within a multimodal context.

While prior research often utilised the standard Wav2Vec 2.0 model, sometimes without any fine-tuning, Chen et al.~\cite{pepino2021emotion} highlighted the necessity for task-specific adjustments. They demonstrated that the Wav2Vec 2.0 model, originally trained for automatic speech recognition (ASR), requires fine-tuning to better align with the nuances of emotion recognition tasks, due to the distinct nature of these applications. Similarly, Pepino et al.~\cite{chen2023exploring} explored the integration of additional audio features, such as eGeMAPS~\cite{7160715}, with the Wav2Vec 2.0 framework. This combination offers a structured approach to enhance the model's performance by providing supplementary contextual cues for emotion analysis.

Building upon similar research, Wagner et al.~\cite{wagner_2022_6221127, wagner2023dawn} conducted a comprehensive analysis across various tasks and models, revealing that extra fine-tuning aimed at automatic speech recognition (ASR) fails to enhance performance. This finding implies that the default Wav2Vec 2.0 model might not be optimally configured for emotion recognition tasks. Accordingly, their study highlights the importance of domain-specific training to better prepare the model for emotion detection.

Recent advancements in affective computing reveal that audio features 
are pivotal for recognising emotions, with multimodal integration 
highlighting the complexity and computational demands of current models.
The use of advanced fusion techniques to combine diverse data 
underscores the challenge of balancing sophisticated analysis with the 
high computational needs of these complex architectures.

\section{Dataset and Challenge}
\subsection{Challenge}
The dataset~\cite{kollias20246th} for the Emotional Mimicry Intensity Challenge (EMI-Challenge) includes over 30 hours of audiovisual content from 557 participants, recorded in natural settings using webcams. Participants were prompted to mimick seed videos showing a person expressing a particular emotion. After that, they had to rate the intensity of the resulting emotional experience on a 0-100 scale. Participants evaluated videos displaying emotional mimicry without knowing the intended emotion to be mimicked, ensuring their judgements remained unbiased. The dataset is split into training (8072 videos, approx. 15 hours), validation (4588 videos, approx. 9 hours), and test set (4586 videos, approx. 9 hours) without speaker overlap. Training and validation sets come with annotations, while test set predictions are submitted for evaluation. 
The dataset includes detected faces at 6 fps, features from Vision Transformer (ViT)~\cite{caron2021emerging, dosovitskiy2020image} for video and Wav2Vec 2.0 features~\cite{baevski2020wav2vec} and the raw videos and audios. Six different emotional expressions are annotated, namely: ``Admiration'', ``Amusement'', ``Determination'', ``Empathic Pain'', ``Excitement'', and ``Joy''.

The performance on the respective data-split ($\rho_\mathit{VAL}$, $\rho_\mathit{TEST}$) as reported in \Cref{tab:eval} and \Cref{tab:archchoice} in this task is measured with the Pearson’s Correlation Coefficient $\rho \in [-1, 1]$ averaged over all predicted emotions:

\begin{equation}
    \rho = \frac{1}{6}\sum_{i=1}^6{\rho_i} \,,
\end{equation}

\noindent with $\rho_i$ as

\begin{equation}
    \rho_i = \frac{Cov(X_{i, pred}, Y_{i, label})}{\sqrt{Var(X_{i, pred})}\sqrt{Var(Y_{i, label})}} \, .
    \label{eq:correlation}
\end{equation}

\begin{figure}[h!]
    \centering
    \includegraphics[width=\linewidth]{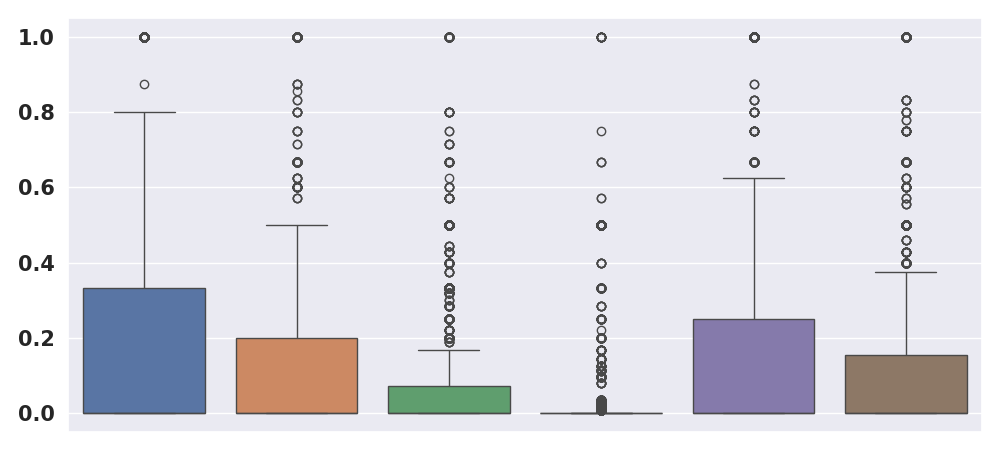}
    \includegraphics[width=\linewidth]{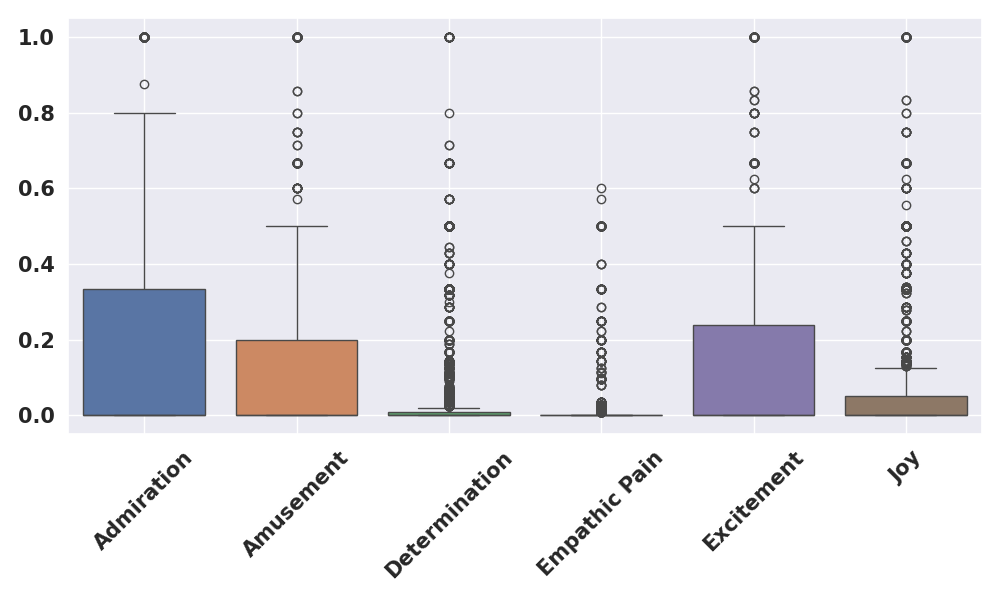}
    \caption{\textbf{Boxplots of the label distribution:} train (upper plot) and validation (lower plot).}
    \label{fig:boxplots}
\end{figure}

\noindent In the boxplots depicted in \Cref{fig:boxplots} for both the training and validation datasets, we observe a significant imbalance in the distribution of regression targets. The distribution for validation also mainly differs from training in the classes ``Determination'' and ``Joy''. 

\subsection{Descriptive Data Analysis} \label{sec:dataset}

The data in both plots are heavily skewed, with a vast majority of values congregating near zero. This skewness highlights the relative rarity of regression targets assigned the value '1', setting them apart as the less frequent outcomes amidst a dominant backdrop of values near zero - accounting for $55.3\%$ to $89.9\%$ of labels for training, respectively $34.1\%$ to $52.4\%$ for validation. Such a distribution presents challenges for learning algorithms, particularly in terms of accurately predicting or learning from these rare extreme values. 

Audios are of constant quality - mono, $16\mathit{kHz}$ sampling rate, and $64 \mathit{kb/s}$ bitrate.

\begin{figure}[]
    \centering
    \includegraphics[width=\linewidth]{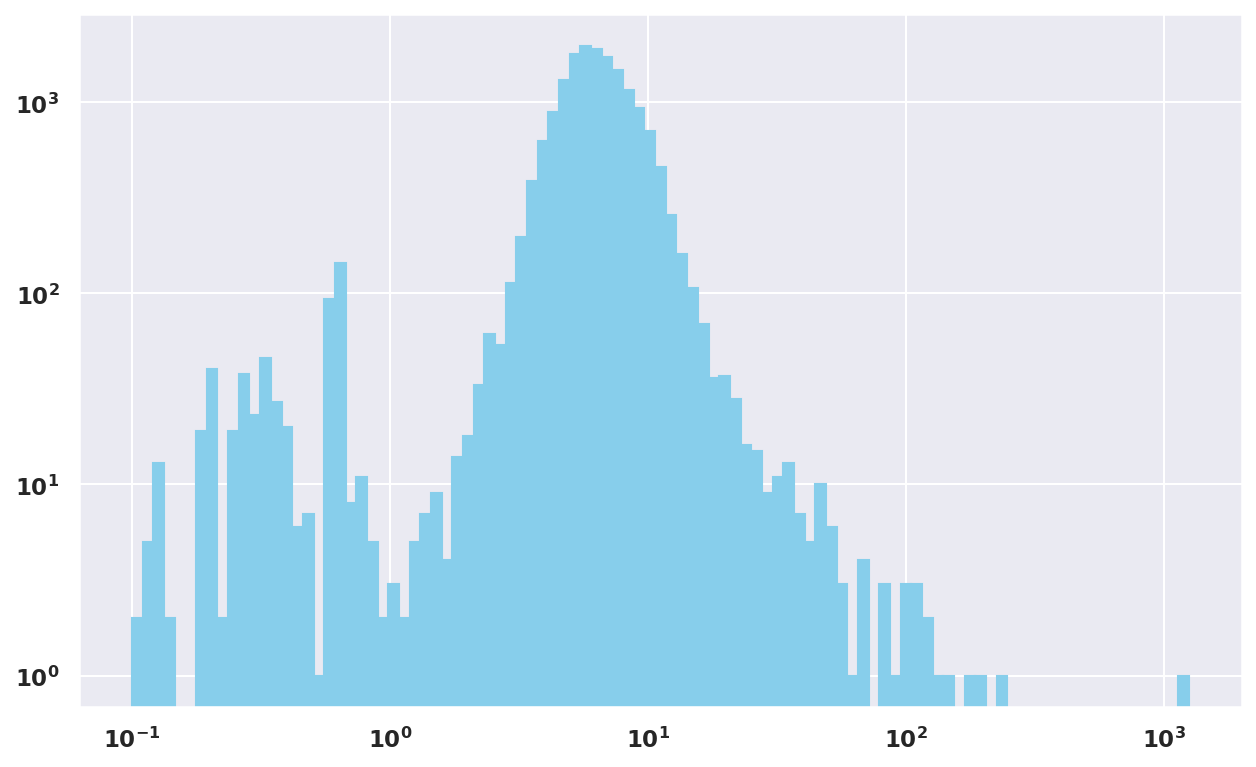}
    \\\vspace{0.5em}
    \includegraphics[width=\linewidth]{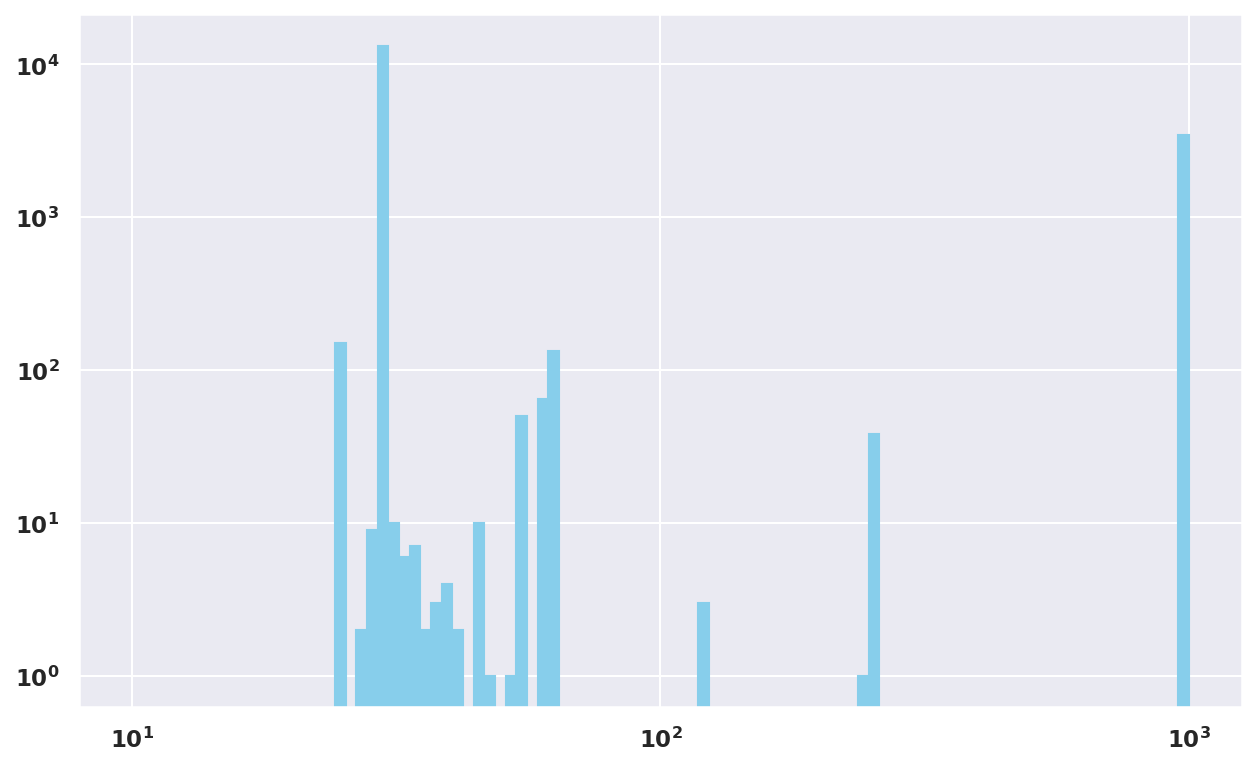}
    \caption{\textbf{Video duration's in seconds (top) and fps (bottom) distribution over all data splits in log-log scale.}}
    \label{fig:fpsdistribution}
\end{figure}

\Cref{fig:fpsdistribution} highlights a key challenge in video analysis: the variability in video lengths and recording diversity, which impacts the stability of frames per second (fps) and significantly increases memory demands for batch processing videos of different lengths, or even for videos of same length with highly different fps. This variability introduces complexities for deep learning models, which rely on consistent input data formats and sizes for optimal training and inference.

For the main part of video lengths, we see a Gaussian distribution, with the left tail standing out. The distribution ranges from $0.1s$ to $1249.9s$ with $6.93s$ on average. $533$ ($3.1\%$) are shorter than $1s$, $753$ ($4.4\%$) longer than $12s$. Former are likely an issue of data quality, as mimicking an emotion in less than $1s$ seems infeasible, or rather is the result of recording errors. The latter have to be considered for a trade-off between batch processing, finite VRAM, and seeing the whole length of the videos. We can safely say, that the video with a duration longer than $2$~hours is either a recording error or mimics the same emotion for a very long time. The remaining $752$ long videos have to be considered for the trade-off -- we cut videos to a length of maximum $12s$, thereby seeing $95.6\%$ of videos wholly, and the cut data amounts to $1.4\%$ of total data -- a good compromise.

The fps distribution is unexpected - since the recordings happen in a domestic webcam-based scenario, you would expect to see $2$ to $3$ bars -- $25, 30$ and maybe $60$ fps. The values range from $25$ to $1000$ fps, with an average of $226$. $62.6\%$ of videos are recorded at $30$ fps, $20.1\%$ at $1000$ fps. There are also exotic values present like $29.83$ or $62.5$ fps. This means, that frames of different videos resembles a highly variable different amount of time, which affects audio alignment to said frames. Extracting faces at a fixed fps does not alleviate this completely, as you have to extract at a higher rate than the target fps, since the face detector can fail, thereby retaining some of the variance of time resembled per frame. \Eg for video ``11437'', originally recorded at $30$ fps and a duration of $6.54s$, you expect $6.54s \cdot 6\mathit{fps} = 39$ frames, but are given $42$ frames, resembling a frame rate of $42 / 6.54 = 6.42$. Analogously, for ``15709'' this results in $5.84\mathit{fps}$.  

\begin{figure}[]
    \centering
    \begin{subfigure}{0.49\linewidth}
        \centering
        \includegraphics[width=\linewidth,left]{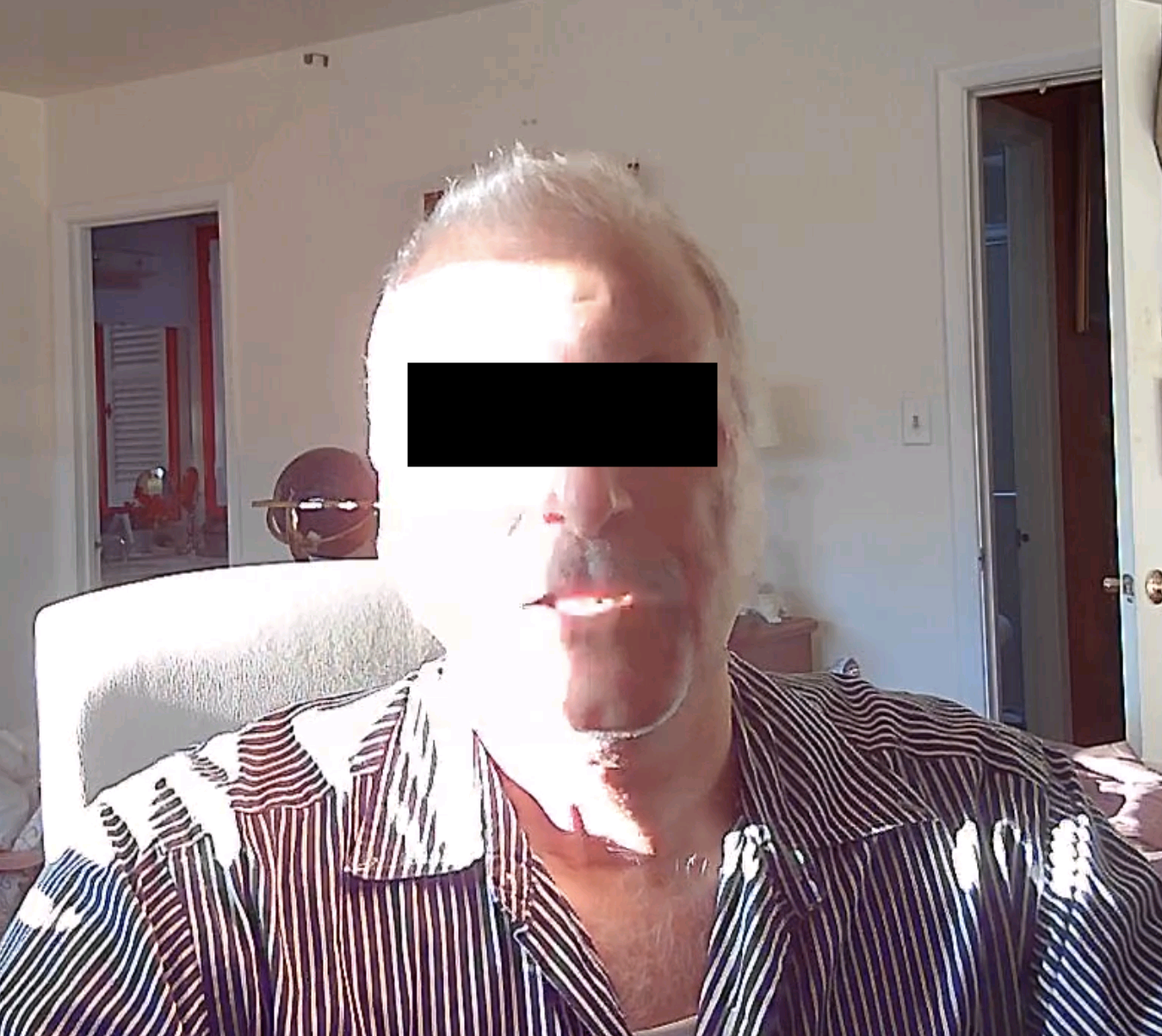}    
    \end{subfigure}
    \begin{subfigure}{0.435\linewidth}
        \centering
        \includegraphics[width=\linewidth,right]{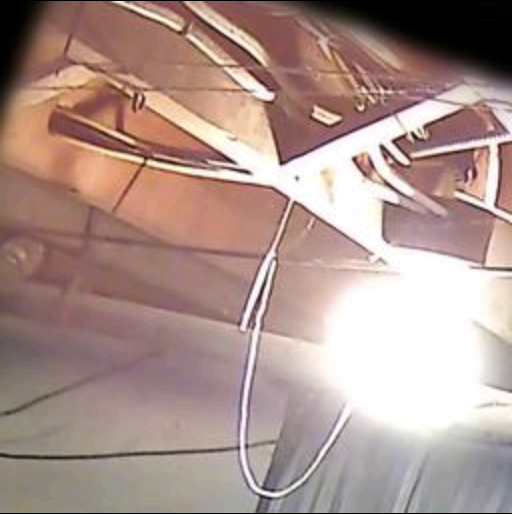}    
    \end{subfigure}
    
    \caption{\textbf{Example of particular challenging videos:} overall quality, \eg illumination (left, manual crop), affects downstream tasks, \eg face detection (right).}
    \label{fig:challengingvideo}
\end{figure}

\Cref{fig:challengingvideo} presents one of those issues where crucial facial features are occluded, adding another layer of complexity, failing downstream tasks. For deep learning algorithms, particularly those focused on facial recognition or emotion detection, occlusions can severely hinder the model's ability to accurately identify and analyse key features. This obstruction challenges the model's learning process as well as the above aligning frames to audio issue, which is why we focus on an audio-only approach.

\section{Methodology} 
In our methodology, code publicly available here\footnote{\url{https://github.com/Skyy93/CVPR2024_abaw}},
we harness the capabilities of the pre-trained Wav2Vec 2.0 model~\cite{wagner_2022_6221127, wagner2023dawn} as a core feature extractor. This choice is underpinned by Wav2Vec 2.0's proven efficiency in distilling important acoustic features, particularly for emotional speech analysis, a strength validated by numerous studies~\cite{guofeng2023llm, groz2023electra, ding2023multimodal}. Our decision to rely solely on audio data is informed by the practical challenges encountered in therapeutic and real-world scenarios, such as the potential occlusion of facial expressions, which can compromise the reliability of visual cues. Moreover, while incorporating multimodal data can enrich the analysis, it invariably escalates computational demands.

\begin{figure}
\begin{center}
     \includegraphics[width=\linewidth]{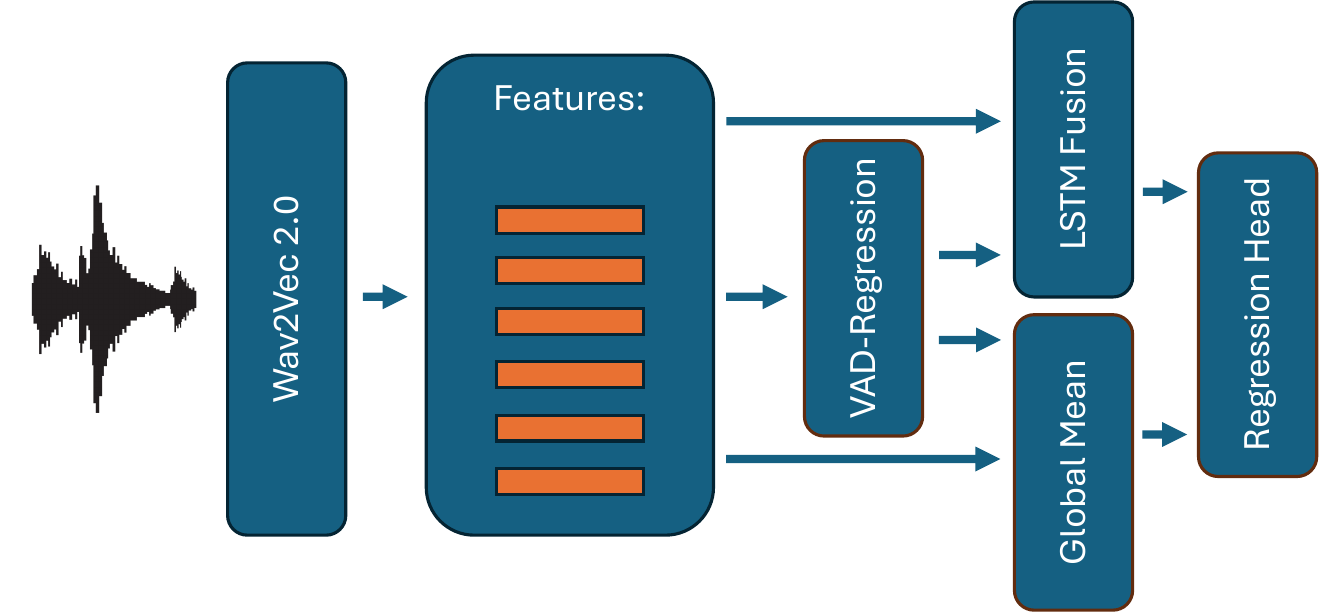}
    \end{center}
      \caption{\textbf{Architecture overview of our approach.} We use a pre-trained Wav2Vec 2.0 model~\cite{wagner_2022_6221127} with a Valence-Arousal-Dominance (VAD) module and extract the features as well as the VAD predictions. To leverage global context we use a global vector and fuse the temporal features in an LSTM.}
      \label{fig:method}
\end{figure}

To augment the model’s ability to predict emotional mimicry intensity, 
which includes nuanced expressions like ``Admiration'', ``Amusement'', ``Determination'', ``Empathic Pain'', ``Excitement'', and ``Joy'', we employ a unimodal multi-task fusion approach by incorporating a 
Valence-Arousal-Dominance (VAD) prediction module, as shown in~\Cref{fig:method}. The inclusion of VAD is strategic, aimed at embedding an additional layer of emotional granularity. Valence captures the positivity or negativity of an emotion, arousal reflects the intensity of emotional activation, and dominance denotes the control level within the emotional experience. By integrating these dimensions, we aim to encapsulate a more comprehensive emotional spectrum, thereby providing a richer contextual basis for predicting specific emotional mimicry intensities.



\paragraph{End-to-end Description}
The raw mono-channel audios, sampled at $16\mathit{kHz}$, are cut to a maximum length of $12s$, \cf~\Cref{sec:dataset}, and prepared by a Wav2Vec2.0-preprocessor. These preprocessed, variable length audios are then batch-processed by a Wav2Vec2.0 model, fine-tuned to VAD regression~\cite{wagner2023dawn}, having continuous VAD values ranging from 0 to 1. The sequence lengths are reduced by the model to maximum of 599, for an initial maximum length of $12s \cdot 16\mathit{kHz} = 192000$. Both the 1024-dimensional model features, as well as the 3-dimensional VAD regressions, are fed concatenated as 1027 dimensions into mean pooling and an LSTM. Both eliminate the sequence dimension, with the former outputting a per audio global mean, and the latter yielding temporal aware features. These $2 \cdot 1027 = 2054$ features are then concatenated and finally processed by a regression head, consisting of a dense layer with tanh activation, and a projection layer, outputting the final 6 dimensional emotional mimicry intensities. For the architectural choice of our regression head we follow the VAD module from~\cite{wagner2023dawn}.

\paragraph{Implementation Details} In our method we unfreeze the Wav2Vec 2.0 model together with the pre-trained VAD module. To improve the generalisation of the LSTM a dropout of 0.1 serves as additional regularisation. We adopt a learning rate of 1e-4, applying cosine decay. For our loss function, we use Mean Squared Error (MSE) because initial experiments with Concordance Correlation Coefficient (CCC) loss and Pearson Correlation Loss did not yield performance improvements. The training process spans 30 epochs, incorporating early stopping with a batch size of 32.

\section{Evaluation}
As demonstrated in ~\Cref{tab:eval}, the proposed methodology surpasses the established baseline and secures the position of runner-up in the 6th Workshop and Competition on Affective Behavior Analysis in-the-wild. The baseline consists of pre-extracted Wav2Vec 2.0 features with a linear layer for the auditory modality. The vision features are extracted with a ViT and processed by a 3-layer gated recurrent unit (GRU) network. For the multimodal fusion the predictions were averaged.
A detailed analysis of the architectural decisions is given in \Cref{sec:archchoice}. Initially, the study explored the use of vision input as a foundational benchmark. However, due to limitations and challenges mentioned in~\Cref{sec:dataset} and~\Cref{sec:explofinding}, this approach is subsequently discontinued. Notably, incorporating vision input was found to detrimentally affect the performance of our unimodal model, particularly when subjected to joint fine-tuning.

For the test set, additional regularisation is applied through the incorporation of a $10\%$ Dropout in the LSTM layer. The highest scores are achieved by training the model on a combined dataset of training and validation sets, thereby enhancing the model's capacity to address underrepresented emotions. 

While our approach does not leverage the vision data, Savchenko et al.~\cite{savchenko2024hsemotion4} use a variety of different Convolutional Neural Networks (CNNs) to encode the facial information of the dataset. The facial descriptor models are note fine-tuned and a simple linear layer is optimised on the statistical feature descriptor vector of the whole video. Wav2Vec 2.0 is also used for audio processing in their approach, albeit without fine-tuning or additional pre-training.

Yu et al.~\cite{yu2024efficient3}  employ vision features extracted by a ResNet-18, pre-trained on the AffectNet database~\cite{mollahosseini2017affectnet}, in conjunction with predicted Action Units and Wav2Vec 2.0 features. Considering the temporal characteristics of the video data, a Temporal Convolutional Network (TCN) is utilised for additional feature refinement across both modalities. To address long-range dependencies in the vision data, a Transformer encoder is applied to the vision features processed by the TCN. Consistent with the above research, it is observed that vision features underperform in comparison to audio features. To fuse the predictions of both modalities a late fusion strategy is employed, which provide additional performance gain, unlike our work.

A new feature extractor specifically designed for facial feature extraction was introduced by Zhang et al.~\cite{zhang2024affective1}, utilising a Masked Autoencoder~\cite{he2022masked} (MAE) trained over 800 epochs on an extensive composite dataset. This dataset, integrates multiple sources - Affect-Net~\cite{mollahosseini2017affectnet}, CASIA-WebFace~\cite{yi2014learning}, CelebA~\cite{liu2015deep}, IMDB-WIKI~\cite{Rothe-IJCV-2018}, and WebFace260M~\cite{zhu2021webface260m} - resulting in an expansive collection of 262 million images. Beyond facial analysis, their approach integrates audio encoding through VGGish~\cite{chen2020vggsound} and textual feature extraction. The methodology utilises Transformer encoders tailored to each data modality, merging the outcomes via a voting strategy. This ensemble technique results in superior performance, albeit with the trade-off of substantial initial training requirements.

Across the compared approaches, a common finding is the underperformance of vision encoders compared to audio modalities, with sound consistently emerging as the most indicative modality for emotional analysis. Our methodology, distinct in its exclusive reliance on audio without the integration of vision or textual data, offers an efficient and streamlined approach for inferring the intensity of emotional mimicry, standing out for its simplicity.
\begin{table}
    \centering
   \begin{center}
    \resizebox{\columnwidth}{!}{ 
    \begin{tabular}{l|cc|cc} \hline \hline
       Model   & Vision & Audio & $\rho_\mathit{VAL}$ & $\rho_\mathit{TEST}$\\ \hline
        Baseline~\cite{kollias20246th}   & X & - & - & .090\\
        Baseline~\cite{kollias20246th}  & - & X & - & .240\\
        Baseline~\cite{kollias20246th}  & X & X   & - & .250\\
        Savchenko et al.~\cite{savchenko2024hsemotion4} & X & X  & .289 & .331\\
        Yu et al.~\cite{yu2024efficient3} & X & X  & .328 & .359\\
        Zhang et al.~\cite{zhang2024affective1} & X & X  & \textbf{.463} & \textbf{.718}\\
        \hline
        Ours\textsubscript{train}    & X & - & .013 & -  \\
        Ours\textsubscript{train}    & X & X & .198 & -  \\
        Ours\textsubscript{train} (freezed)   & - & X & .262 & -  \\
        Ours\textsubscript{train}    & - & X & .386 & .461  \\
        Ours\textsubscript{train} (w/ Dropout)  & - & X & \underline{.389} & .465  \\
        Ours\textsubscript{train+val}    & - & X & - & .522  \\
        Ours\textsubscript{train+val} (longer Training)   & - & X & - & \underline{.554}  \\
     \hline \hline
    \end{tabular}
    }
    \end{center}
        \caption{\textbf{Quantitative comparison of our approach.} } We compare our solution with the Top-3 other solutions and achieve second place in the EMI Challenge.
    \label{tab:eval}
\end{table}
\section{Ablation}
\label{sec:ablation}
Our study adopts a unimodal approach, centering on the auditory modality as the primary source of input. This decision is underscored by the employment of the Wav2Vec 2.0 framework, renowned for its ability to capture rich, nuanced acoustic features from raw audio data. In this ablation we feature multiple experiments over our architectural choice and the challenges that arise when using a multimodal approach. 
\subsection{Architectural Choices}
\label{sec:archchoice}
In our study, we systematically evaluated various configurations of the Wav2Vec 2.0-large model. Our results, detailed in~\Cref{tab:archchoice}, demonstrate the incremental impact of different architectural elements on model performance, measured by the correlation coefficient ($\rho_\mathit{VAL}$).

The foundational experiment utilised the fine-tuning of a Wav2Vec 2.0-large model~\cite{baevski2020wav2vec} and its multilingual derivate~\cite{conneau2020unsupervised}, yielding a $\rho_\mathit{VAL}$ of $0.017$ and $0.021$. This initial outcome highlighted the limitations of using models pre-trained on standard Automatic Speech Recognition (ASR) tasks, such as those involving the LibriSpeech dataset~\cite{7178964}, for complex emotional recognition tasks.

Subsequent experiments involved incrementally adding components to the model architecture, such as Global Vector, Regression Head, and LSTM layers. The introduction of a Regression Head and Global Vector improved the $\rho_\mathit{VAL}$ to $0.356$, indicating the significance of incorporating model components that enhance the representation of speech features relevant to emotion recognition.

Further incorporation of LSTM layers, known for their effectiveness in capturing temporal dependencies in data, resulted in a $\rho_\mathit{VAL}$ of $0.375$. A notable point from our findings is that neither the global vector nor the VAD (Valence, Arousal, Dominance) module significantly enhances performance on their own. However, when integrated, they collectively offer an additional enhancement.

The most comprehensive model configuration, which included all evaluated components (Global Vector, Regression Head, LSTM, and VAD Head), achieved the highest $\rho_\mathit{VAL}$ of $0.386$. 

Our experimental results underscore the critical role of tailored pre-training and architectural design in enhancing model performance on specialised tasks like estimating the emotion mimicry intensity. It is evident that generic ASR pre-training is insufficient for complex tasks requiring nuanced emotional understanding, thereby highlighting the need for domain-specific adaptations in model training and architecture.
\begin{table}
    \centering
   \begin{center}
    \resizebox{\linewidth}{!}{ 
    \begin{tabular}{l|cccc|cc} \hline \hline
       Model  & Global Vector & Reg. Head & LSTM  & VAD Head  & $\rho_\mathit{VAL}$ \\ \hline
        W2V2-L~\cite{baevski2020wav2vec}  & X & - & - & - & $.017$   \\
        W2V2-L XLSR~\cite{conneau2020unsupervised}   & X & - & - & - & $.021$   \\
        W2V2-L Audeering~\cite{wagner_2022_6221127}   & X & - & - & - & $.342$   \\
        W2V2-L Audeering~\cite{wagner_2022_6221127}   & X & X & - & - & $.356$   \\
        W2V2-L Audeering~\cite{wagner_2022_6221127}   & - & X & X & - & $.375$   \\
        W2V2-L Audeering~\cite{wagner_2022_6221127}   & X & X & X & - & $.377$   \\
        W2V2-L Audeering~\cite{wagner_2022_6221127}   & - & X & X & X & $.375$   \\
        W2V2-L Audeering~\cite{wagner_2022_6221127}   & X & X & X & X & \textbf{$.386$}   \\
     \hline \hline
    \end{tabular}
    }
    \end{center}
    \caption{\textbf{Comparison of our design choices.}}
    \label{tab:archchoice}
\end{table}

\subsection{Exploratory Findings}
\label{sec:explofinding}

Zero-padding, \ie filling missing video data with black images or audio with silence to create fixed length batch tensors, was not only detrimental to memory efficiency, thereby batch size and training duration, but also to training itself -- the padding amounted to $34\%$ of batch data on average, often disallowing the model to learn anything at all. $\rho$ in the region of $-0.1$ was not uncommon.

Correlation as loss instead of MSE, \ie

\begin{equation}
    L = 1 - \rho \,, \, L \in [0, 2] \,,
\end{equation}

\noindent did not work aswell. With the huge imbalance in the labels towards a single value, \cf~\Cref{sec:dataset}, the models quickly learn to make a constant prediction, thereby having an undefined variance (\cf~ \Cref{eq:correlation}), therefore having an undefined loss, blocking training via backpropagation completely.

Fine-tuning a generally pretrained vision model, \eg Convnext~\cite{liu2022convnet} or DinoV2~\cite{oquab2024dinov2}, yielded some improvement for $\rho$ from $-0.10$ to $-0.01$, but remained underwhelming to a degree, that we doubted the soundness of our pipeline. To check, we trained and validated using a 2:1 split both on official training and official validation split -- in both cases $\rho$ rose to about $0.18$. This lead us to the assumption, that the pipeline itself works, but that the vision models do not generalise at all from training to validation split, but immediately overfit to the data.

To mitigate the variance of the time between face detection frames, \cf~\Cref{sec:dataset}, we tried face detection using BlazeFace~\cite{bazarevsky2019blazeface} at a constant framerate. Also increasing the crop area from 160x160 to 256x256, and aligning the detected faces afterwards. But this did not change the performance of beforehand vision models noticeably.

We then replaced the vision models with models trained for facial expression recognition, \eg LibreFace~\cite{chang2023libreface} and EfficientNet~\cite{efficientnet}. The performance did somewhat improve to our best vision-only performance of $\rho = 0.013$, respectively multimodal of $\rho = 0.198$, but they still disappoint.

This then lead us to dropping the vision modality completely, yielding a $32\%$ boost to the performance ($\rho = 0.262$) when training on audio only.

\section{Discussion}
In our study, we concentrated on analysing emotional mimicry intensity through audio data, moving away from the common focus on facial expressions in affective computing. Despite having a comprehensive multimodal dataset, we found that adding facial images to the analysis decreased its effectiveness, as shown by lower Pearson correlation coefficients when including vision compared to audio-only results. Our findings emphasise the need to choose the right modality for emotional analysis and hint at audio's unique potential in this field. Notably, our approach stood out among challenge participants by employing a unimodal strategy, prioritising computational speed and resource efficiency. This distinctive choice underlines the potential of streamlined, audio-focused analysis in settings where efficiency is paramount. Additionally, an intriguing avenue for future research could be the development of multimodal models that not only process audio signals but also interpret the textual content within these signals, enriching the emotional analysis. Further research could explore aligning audio with facial expressions, addressing the challenging task of effectively integrating these modalities.
\section{Conclusion}
Our study utilised the pre-trained Wav2Vec 2.0 model for emotional speech analysis, focusing on audio data to address real-world challenges like facial occlusion. We enhanced emotion detection by integrating a Valence-Arousal-Dominance (VAD) module with our model, aiming for a deeper emotional understanding. Our multi-task approach, validated through an extensive ablation study, showed promising results in predicting emotional mimicry intensities. This work, which achieved second place in the Emotional Mimicry Challenge at the 6th Affective Behavior Analysis in-the-wild Workshop, highlights the potential of audio-focused analysis in affective computing.
\section*{Acknowledgement}
The authors gratefully acknowledge the computing time granted by the Institute for Distributed Intelligent Systems and provided on the GPU cluster Monacum One at the University of the Bundeswehr Munich.

This work was partially funded by the KodiLL project (FBM2020, Stiftung Innovation in der Hochschullehre).



{
    \small
    \bibliographystyle{ieeenat_fullname}
    \bibliography{main,arxiv/egbib}
}


\end{document}